\documentclass[amsmath,amssymb,a4paper]{revtex4}

\usepackage{bm}
\usepackage[mathscr]{eucal}
\def\II{I\hspace{-.1em}I}

\def\mp{m_p}
\def\me{m_e}
\def\mH{m_H}

\def\np{n_p}
\def\ne{n_e}
\def\nH{n_H}
\def\nb{n_b}
\def\v{\vec{v}}
\def\vp{\vec{v}_p}
\def\ve{\vec{v}_e}
\def\vH{\vec{v}_H}
\def\vb{\vec{v}_b}
\def\dnpe{\delta n_{pe}}
\def\dv{\delta \vec{v}}
\def\dvpe{\delta \vec{v}_{pe}}
\def\dvpH{\delta \vec{v}_{pH}}
\def\dveH{\delta \vec{v}_{eH}}
\def\dvbH{\delta \vec{v}_{bH}}

\def\dvpeI{\delta \vec{v}_{pe}^{(I)}}

\def\dvbHI{\delta \vec{v}_{bH}^{(I)}}
\def\dvpeII{\delta \vec{v}_{pe}^{(\II)}}

\def\dvbHII{\delta \vec{v}_{bH}^{(\II)}}

\def\E{\vec{E}}
\def\EI{\vec{E}^{(I)}}
\def\EII{\vec{E}^{(\II)}}
\def\B{\vec{B}}
\def\BI{\vec{B}^{(I)}}
\def\BII{\vec{B}^{(\II)}}

\def\dt{\partial_t}

\begin{document}

\title{Behavior of magnetic fields around the cosmic recombination}

\author{%
Keitaro Takahashi \footnote{E-mail address:keitaro@yukawa.kyoto-u.ac.jp}}
\affiliation{%
Yukawa Institute for Theoretical Physics, Kyoto University,
Kyoto 606-8502, Japan}

\date{\today}

\begin{abstract}
Several mechanisms have been proposed to generate primordial magnetic fields
and it is often assumed that magnetic fields are not affected by a sharp drop
in ionization rate due to the cosmic recombination. We investigate the validity of
the assumption by studying the behavior of magnetic fields and fluid motion
around recombination. Fluid equations including the effect of recombination
are considered for protons, electrons and neutral hydrogens separately, combining
Maxwell equations. We find that the residual ionization rate required for
the conservation of magnetic field at cosmological scales is about $10^{-10}$,
which is much smaller than the standard value $\sim 10^{-4}$. Further we will
show the acceleration of protons and electrons in the process of recombination
which conpensates the decrease in the carrier of electric current in order to
preserve electric current and then magnetic fields.
\end{abstract}

\maketitle

\section{Introduction}

There have been a lot of studies on cosmological magnetogenesis
before recombination. Among them are generation from nonlinear cosmological
perturbations during radiation-dominated era
\cite{Hogan00,Dolgov04,Matarrese05,GopalSethi05,KT1,KT2,SiegelFry06,KT3,TCA,KT4,Hollenstein07},
phase transition \cite{QuashnockLoebSpergel89,BaymBodekerMcLerran96} and breaking
of conformal invariance during inflationary epoch
\cite{TurnerWidrow88,Ratra92,BambaYokoyama04,BambaSasaki07,Bamba07}
(see \cite{GrassoRubinstein01,Widrow03,Giovannini04} for comprehensive reviews).
One of their motivations is to give seed fields to explain the origin of
galactic magnetic fields through the dynamo amplification.
Also, primordial magnetic fields are expected to play a important
role in structure formation. In this context, it is usually assumed
that magnetic fields survive through cosmic recombination.
This assumption seems valid because, even though the ionization rate drops to
about $10^{-4}$ \cite{WongMossScott07} after recombination, magnetohydrodynamic
approximation is expected to be still valid due to the residual charged particles
and comoving magnetic fields are then conserved.

Because this is a crucial assumption in the studies on magnetic fields
in the early universe, its validity is worth investigating in detail and this is
the subject of this paper. We will estimate how small the residual
ionization rate could be in order to preserve magnetic fields with
cosmological scales. Furthermore, the evolution of the relative velocity
between protons and electrons will also be studied in detail. The acceleration
of the relative velocity should accompany the recombination because the conservation
of magnetic fields means the conservation of electric current and then,
noting the decrease in the current carriers, increase in the relative velocity.
Thus, to see this, we need to follow the evolution of protons and electrons
separately. We pursue these problems by solving the equations for three fluids,
protons, electrons and neutral hydrogens, including the effect of recombination and
combining Maxwell equations. Because we are interested in an epoch around
the recombination, the density fluctuations and velocities of the fluids
are so small that linear perturbation would be a very good approximation.
Further we adopt tight coupling approximation \cite{PeeblesYu70} between
the three fluids which is valid when the collision timescale is much
smaller than the dynamical timescale. To make the problem tractable,
the effects of photons will be neglected as is often done in the previous
studies on magnetic field evolution after recombination
\cite{BanerjeeJedamzik04,SethiSubramanian05}. 

In \cite{BanerjeeJedamzik04}, the evolution of primordial magnetic fields
after recombination was considered taking ambipolar diffusion into account
(see also \cite{JedamzikKatalinicOlinto98,SethiSubramanian05,TashiroSugiyama06}).
Because neutral hydrogens do not feel Lorentz force, the presence of magnetic fields
leads to the difference of motion between neutral hydrogens and charged particles.
Then neutral hydrogens act as effective electric resistivity which results
in the diffusion of magnetic fields. This is a nonlinear effect and, therefore,
will not be included in the present analysis. However, it was shown that ambipolar
diffusion is not effective in the tight coupling regime under consideration.
As we will see, we have another type of effective electric resistivity
which appears even at the linear order.

This paper is organized as follows. In section \ref{section:basic},
we present basic equations of our analysis: continuity equations,
Euler equations and Maxwell equations. We will neglect general
relativistic effects including cosmological expansion because
we are interested in electromagnetic phenomena deep inside the horizon
and gravitation would have minor effects there. Instead, we simply adopt
Newtonian gravitational potential which should be determined by the distribution
of cold dark matter, but our final results do not involve gravitational potential
so that we do not need to mention cold dark matter and Poisson equation.
The basic equations are solved in section \ref{section:TCA} using
tight coupling approximation. In order to see the acceleration of protons
and electrons separately, we need to go on to the next to the lowest order.
FInally, we will give discussion and summary in section \ref{section:discussion}.

\section{Basic Equations \label{section:basic}}

The continuity equations of number densities are expressed as,
\begin{eqnarray}
&& \dt \np + \nabla \cdot ( \np \vp ) = - F, \label{conserve-p} \\
&& \dt \ne + \nabla \cdot ( \ne \ve ) = - F, \label{conserve-e} \\
&& \dt \nH + \nabla \cdot ( \nH \vH ) = F, \label{conserve-H}
\end{eqnarray}
where $n_{\alpha}$ and $\v_{\alpha}$ $(\alpha = p,e,H)$ are the number
density and velocity of component $\alpha$, respectively. Here the $F$ terms
represent the rate of recombination and we see that $(\np + \nH)$ is
conserved with a mean value,
\begin{equation}
\np + \nH \approx 3 \times 10^{2} ~ {\rm cm}^{-3} \left( \frac{1+z}{10^3} \right)^3.
\end{equation}
With the same $F$, the momentum continuity equation of each component
can be written as,
\begin{eqnarray}
&& \dt ( \mp \np \vp ) + \nabla \cdot ( \mp \np \vp \vp )
   = e \np ( \E + \vp \times \B ) - e^2 \np \ne \eta \dvpe - \mp \np \nu_p \dvpH
     - \mp \np \nabla \Phi - F \mp \vp, \\
&& \dt ( \me \ne \ve ) + \nabla \cdot ( \me \ne \ve \ve )
   = - e \ne ( \E + \ve \times \B ) + e^2 \np \ne \eta \dvpe - \me \ne \nu_e \dveH
     - \me \ne \nabla \Phi - F \me \ve, \\
&& \dt ( \mH \nH \vH ) + \nabla \cdot ( \mH \nH \vH \vH )
   = \mp \np \nu_p \dvpH + \me \ne \nu_e \dveH
     - \mH \nH \nabla \Phi + F ( \mp \vp + \me \ve ),
\end{eqnarray}
where $m_{\alpha}$ is mass of a particle $\alpha$, $e$ is electric charge,
$\E$ and $\B$ are electromagnetic fields, $\dv_{\alpha \beta} = \v_{\alpha} - \v_{\beta}$
is the velocity difference between components $\alpha$ and $\beta$, and $\Phi$ is
gravitational potential. The electric resistivity, $\eta$, and the collision
frequency between the component $\alpha$ and neutral hydrogen, $\nu_{\alpha}$, are
\cite{Draine83},
\begin{eqnarray}
&& \eta = \frac{\pi e^{2} \sqrt{m_{e}}}{T^{3/2}} \ln{\Lambda}
        \approx 10^{-12} ~ {\rm sec} \left( \frac{1+z}{10^3} \right)^{-3/2}
                \left( \frac{\ln{\Lambda}}{10} \right) \\
&& \nu_e = \langle v^{\rm th}_e \sigma_{eH} \rangle \nH
         \approx 10^{-5} x_H ~ {\rm sec}^{-1}
                 \left( \frac{1+z}{10^3} \right)^{4}, \label{nu-e} \\
&& \nu_p = \langle v^{\rm th}_p \sigma_{pH} \rangle \nH
         \approx 10^{-6} x_H ~ {\rm sec}^{-1}
                 \left( \frac{1+z}{10^3} \right)^{3}, \label{nu-p}
\end{eqnarray}
where $T$ is the plasma temperature, $\ln \Lambda$ is the Coulomb logarithm,
$v^{\rm th}_{\alpha}$ is the thermal velocity of component $\alpha$,
$\beta = \me/\mp$ and $\sigma_{\alpha H}$ is the scattering cross section
between a particle $\alpha$ and a neutral hydrogen. Here we defined
$x_p = \np / (\np + \nH)$ and $x_H = \nH / (\np + \nH)$. Substituting
the continuity equations of the number densities into the momentum
continuity equations, we obtain,
\begin{eqnarray}
&& \dt \vp + (\vp \cdot \nabla) \vp
   = \frac{e}{\mp} (\E + \vp \times \B) - \frac{e^2 \ne \eta}{\mp} \dvpe
     - \nu_p \dvpH - \nabla \Phi, \label{EOMp-full2} \\
&& \dt \ve + (\ve \cdot \nabla) \ve
   = - \frac{e}{\me} (\E + \ve \times \B) + \frac{e^2 \np \eta}{\me} \dvpe
     - \nu_e \dveH - \nabla \Phi, \label{EOMe-full2} \\
&& \dt \vH + (\vH \cdot \nabla) \vH
   = \frac{\mp \np}{\mH \nH} \nu_p \dvpH + \frac{\me \ne}{\mH \nH} \nu_e \dveH
     - \nabla \Phi + \frac{F}{\mH \nH} ( \mp \vp + \me \ve - \mH \vH ), \label{EOMH-full2}
\end{eqnarray}
It is convenient to rewrite equations using the following quantities,
\begin{eqnarray}
&& \nb = \frac{\mp \np + \me \ne}{\mp + \me}, ~~~ \dnpe = \np - \ne, \\
&& \vb = \frac{\mp \np \vp + \me \ne \ve}{\mp \np + \me \ne}, ~~~
   \dvpe = \vp - \ve,
\end{eqnarray}
and conversely,
\begin{eqnarray}
&& \np = \nb + \frac{\beta}{1+\beta} \dnpe, ~~~
   \ne = \nb - \frac{1}{1+\beta} \dnpe, \\
&& \vp = \vb
         + \left[ \frac{\beta}{1+\beta} - \frac{\beta}{(1+\beta)^2} \frac{\dnpe}{\nb} \right]
           \dvpe, ~~~
   \ve = \vb
         - \left[ \frac{1}{1+\beta} + \frac{\beta}{(1+\beta)^2} \frac{\dnpe}{\nb} \right]
           \dvpe.
\end{eqnarray}
In terms of the new variables, the electric current can be expressed as,
\begin{eqnarray}
&& \rho = e (\np - \ne) = e \dnpe, \\
&& \vec{j}
   = e(\np \vp - \ne \ve)
   = e \left[ \nb \dvpe + \dnpe \vb - \frac{1-\beta}{1+\beta} \dnpe \dvpe
           - \frac{\beta}{(1+\beta)^2} \frac{(\dnpe)^2}{\nb} \dvpe \right].
\end{eqnarray}

We write the recombination rate phenomenologically as $F \equiv f \nb$
where $1/f \lesssim 1/H$ is the effective timescale of recombination and
is assumed to be constant in time here. Actually, $f$ is a function of
several quantities such as the ionization fraction and depends on time,
but the detailed time profiles of $\np$, $\ne$ and $\nH$ are not important
for our analysis. One can see that this form of $F$ does not violate
the charge conservation. In fact, we have, from Eqs. (\ref{conserve-p}) and (\ref{conserve-e}),
\begin{eqnarray}
&& \dt \nb + \nabla \cdot (\nb \vb) = -f \nb, \label{conserve-b} \\
&& \dt \dnpe + \nabla \cdot (\dnpe \vb + \nb \dvpe) = 0.
\end{eqnarray}

We solve the above equations of motion and Maxwell equations,
\begin{eqnarray}
&& \dt \E = \nabla \times \B - \vec{j}, \label{Ampere} \\
&& \dt \B = - \nabla \times \E, \label{Faraday}
\end{eqnarray}
to investigate the behavior of three fluids and electromagnetic fields
around cosmological recombination. To do this, we first expand the physical
quantities according to cosmological perturbation theory. At the zeroth
order, the universe is homogeneous and isotropic so that fluids
do not have peculiar velocities and electromagnetic quantities vanish.
Only the number densities have nonvanishing values and, because
we are neglecting the cosmological expansion, they are constant
in time in the absence of the recombination term. At the first order,
the universe becomes inhomogeneous and anisotropic, and we have peculiar
velocities and electromagnetic quantities as well as the number densities
which are functions of space and time. Actually, for the allowed amplitude
of primordial magnetic fields, $\lesssim 10^{-9} ~ {\rm Gauss}$ (comoving),
by CMB observations \cite{Barrow97,Giovannini06,Yamazaki06,Yamazaki08},
magnetic fields can be treated as perturbations. We consider the equations of motion,
(\ref{EOMp-full2}) - (\ref{EOMH-full2}), and Maxwell equations,
(\ref{Ampere}) and (\ref{Faraday}), up to first order in cosmological
perturbations. We see that in this approximation the advection terms and
the Lorentz force terms of equations of motion are neglected.
It should also be noted that the electric current is approximated as
$\vec{j} \approx e \nb \dvpe$. Then the equations of motion and
Maxwell equations reduce to
\begin{eqnarray}
\dt \vp
&=& \frac{e}{\mp} \E - \frac{e^2 \nb \eta}{\mp} \dvpe - \nu_p \dvpH - \nabla \Phi \nonumber \\
&=& \frac{e}{\mp} \E
    - \left( \frac{e^2 \nb \eta}{\mp} + \frac{\beta}{1+\beta} \nu_p \right) \dvpe
    - \nu_p \dvbH - \nabla \Phi \label{EOMp} \\
\dt \ve
&=& - \frac{e}{\me} \E + \frac{e^2 \nb \eta}{\me} \dvpe - \nu_e \dveH - \nabla \Phi \nonumber \\
&=& - \frac{e}{\me} \E
    + \left( \frac{e^2 \nb \eta}{\me} + \frac{1}{1+\beta} \nu_e \right) \dvpe
    - \nu_e \dvbH - \nabla \Phi \label{EOMe} \\
\dt \vH
&=& \frac{\beta}{(1+\beta)^2} \gamma (\nu_p - \nu_e) \dvpe
    + \gamma \left[ \frac{\nu_p + \beta \nu_e}{1+\beta} + f \right] \dvbH - \nabla \Phi,
    \label{EOMH}
\end{eqnarray}
and
\begin{eqnarray}
&& \dt \E = \nabla \times \B - e \nb \dvpe, \label{Ampere-linear} \\
&& \dt \B = - \nabla \times \E, \label{Faraday-linear}
\end{eqnarray}
respectively. Here $\gamma \equiv \np / \nH$ and evolves from $\infty$ to
$\gamma_{\rm end} \sim 10^{-4}$. It should be noted that because we are
neglecting the second and higher order terms, the coefficients of the velocity terms
are zeroth-order quantities. From Eqs. (\ref{conserve-b}) and (\ref{conserve-H}),
the evolution equations for the zeroth-order number densities are,
\begin{equation}
\frac{\dt \nb}{\nb} = - f, ~~~ \frac{\dt \nH}{\nH} = \gamma f.
\end{equation}
Further we rewrite the equations of motion in terms of velocity differences:
\begin{eqnarray}
&& \dt \vb = - \frac{\beta}{(1+\beta)^2} (\nu_p - \nu_e) \dvpe
             - \frac{\nu_p + \beta \nu_e}{1+\beta} \dvbH - \nabla \Phi, \\
&& \dt \dvpe
   = \frac{(1+\beta) e}{\me} \E
     - \left[ \frac{(1+\beta) e^2 \nb \eta}{\me} + \frac{\beta \nu_p + \nu_e}{1+\beta} \right]
       \dvpe
     - (\nu_p - \nu_e) \dvbH, \label{EOMpe-linear} \\
&& \dt \dvbH
   = - \frac{\beta}{(1+\beta)^2} (1+\gamma) (\nu_p - \nu_e) \dvpe
     - \left[ \frac{\nu_p + \beta \nu_e}{1+\beta} (1+\gamma) + \gamma f \right] \dvbH.
   \label{EOMbH-linear}
\end{eqnarray}
Equations for the velocity differences, (\ref{EOMpe-linear}) and (\ref{EOMbH-linear}),
and Maxwell equations, (\ref{Ampere-linear}) and (\ref{Faraday-linear}), consist
our basic equations to investigate the behavior of magnetic fields and other quantities
around cosmological recombination.

\section{Tight Coupling Approximation \label{section:TCA}}

We solve the equations presented in the previous section using tight coupling
approximation \cite{PeeblesYu70,TCA,KT4}. This approximation is based on the
fact that the different components are tightly coupled through frequent scattering
so that their velocities are almost the same, that is,
\begin{equation}
\v_{\alpha} = \v + \v_{\alpha}^{(I)} + \v_{\alpha}^{(\II)} + \cdots, \label{expansion-v}
\end{equation}
where $\v$ is the common velocity and the others are the deviation from it.
The deviation is suppressed by a small parameter which is typically
the ratio of scattering timescale to dynamical timescale $\tau_{\rm dyn}$.
Here the dynamical timescale can be considered as that of acoustic oscillation,
which is comparable or several orders less than Hubble time,
$\tau_{\rm dyn} \lesssim H^{-1} \approx 2 \times 10^{13} ~ {\rm sec} ((1+z)/10^3)^{-3/2}$.
Thus tight coupling approximation is valid because the timescale of Coulomb scattering,
\begin{equation}
\tau_C = \frac{\me}{e^2 \ne \eta}
       \approx 20 x_p^{-1} ~ {\rm sec} \left( \frac{1+z}{10^3} \right)^{-3/2},
\end{equation}
and those of scatterings involving neutral hydrogens, Eqs. (\ref{nu-e}) and (\ref{nu-p}),
are much smaller than the dynamical timescale. Eq. (\ref{expansion-v}) shows that
the velocity differences start from the first order,
\begin{equation}
\dv_{\alpha \beta}
= \dv_{\alpha \beta}^{(I)} + \dv_{\alpha \beta}^{(\II)} + \cdots \nonumber \\
= \left( \v_{\alpha}^{(I)} - \v_{\beta}^{(I)} \right)
    + \left( \v_{\alpha}^{(\II)} - \v_{\beta}^{(\II)} \right) + \cdots. \\
\end{equation}
Then we see that, in the equations of motion (\ref{EOMpe-linear}) and
(\ref{EOMbH-linear}), the time derivative term is higher order than the collision term.

Electromagnetic fields also start from the first order because the exact tight coupling
of protons and electrons do not allow their existence.
\begin{eqnarray}
\E &=& \E^{(I)} + \E^{(\II)} + \cdots, \\
\B &=& \B^{(I)} + \B^{(\II)} + \cdots.
\end{eqnarray}
We will solve Eqs. (\ref{Ampere-linear}), (\ref{Faraday-linear}), (\ref{EOMpe-linear})
and (\ref{EOMbH-linear}) up to second order in tight coupling approximation.
Here it should be noted that this expansion is independent of the expansion
with respect to cosmological perturbations argued in the previous section.
For further details of tight coupling approximation, see \cite{TCA,KT4}.

\subsection{tight coupling approximation I}

The evolution equations of velocity differences and Maxwell equations at the lowest
order in tight coupling approximation are,
\begin{eqnarray}
&& \frac{(1+\beta) e}{\me} \EI
   - \left[ \frac{(1+\beta) e^2 \nb \eta}{\me} + \frac{\beta \nu_p + \nu_e}{1+\beta} \right]
     \dvpeI
   - (\nu_p - \nu_e) \dvbHI = 0, \label{Ohm-I} \\
&& - \frac{\beta}{(1+\beta)^2} (1+\gamma) (\nu_p - \nu_e) \dvpeI
   - \left[ \frac{\nu_p + \beta \nu_e}{1+\beta} (1+\gamma) + \gamma f \right] \dvbHI
   = 0, \label{dvbH-I} \\
&& \dt \EI = \nabla \times \BI - e \nb \dvpeI, \label{Ampere-I} \\
&& \dt \BI = - \nabla \times \EI. \label{Faraday-I}
\end{eqnarray}
From now on, we neglect the $\gamma f$ term in Eq. (\ref{dvbH-I}) because it is much
smaller than $\nu_p$. We can express $\dvbHI$ and $\EI$ by $\dvpeI$ using
Eqs. (\ref{Ohm-I}) and (\ref{dvbH-I}),
\begin{eqnarray}
&& \dvbHI = - \frac{\beta}{1+\beta} \frac{\nu_p - \nu_e}{\nu_p + \beta \nu_e} \dvpeI,
   \label{dvbH-I-dvpe} \\
&& \EI
   = \left( e \nb \eta + \frac{\me}{e} \frac{\nu_p \nu_e}{\nu_p + \beta \nu_e} \right)
     \dvpeI
     \equiv e \nb \eta_{\rm eff} \dvpeI. \label{E-I-dvpe}
\end{eqnarray}
The latter equation is an effective Ohm's law and the effective electric resistivity is,
\begin{eqnarray}
&& \eta_{\rm eff} = \eta + \frac{\me}{e^2 \nb} \frac{\nu_p \nu_e}{\nu_p + \beta \nu_e}, \\
&& \frac{\me}{e^2 \nb \eta} \frac{\nu_p \nu_e}{\nu_p + \beta \nu_e}
   = \frac{\nu_p \nu_e \tau_C}{\nu_p + \beta \nu_e}
   \approx \nu_e \tau_C
   \approx 2 \times 10^{-4} \gamma^{-1}.
\end{eqnarray}
We can see that the electric resistivity is mainly contributed from Coulomb collision
in the early phase, $\gamma \gg 1$, and collision between charged particles and
neutral hydrogens can dominate in the late phase if $\gamma_{\rm end} < 10^{-4}$.
In the realistic case with $\gamma_{\rm end} \sim 10^{-4}$, the contribution
from neutral hydrogen is at most comparable to that from Coulomb collision.

To solve Eq. (\ref{Ampere-I}), we compare the time derivative of electric field,
$\EI$, and the velocity difference between protons and electrons, $\dvpeI$,
using Eq. (\ref{E-I-dvpe}),
\begin{equation}
\frac{| \dt \EI |}{| e \nb \dvpeI |}
\approx (\tau_{\rm dyn}^{-1} + f) \eta \ll 1
\end{equation}
where we replaced the time derivative with $\tau_{\rm dyn}$. Thus, $\EI$
in Eq. (\ref{Ampere-I}) is higher order and should appear at the next order.
Then, we have approximately,
\begin{equation}
e \nb \dvpeI = \nabla \times \BI, \label{dvpe-I-sol}
\end{equation}
and accordingly, from Eqs. (\ref{dvbH-I-dvpe}) and (\ref{E-I-dvpe}),
\begin{eqnarray}
&& \dvbHI = - \frac{\beta}{1+\beta} \frac{\nu_p - \nu_e}{\nu_p + \beta \nu_e}
              \frac{1}{e \nb} \nabla \times \BI, \label{dvbH-I-sol} \\
&& \EI = \eta_{\rm eff} \nabla \times \BI. \label{E-I-sol}
\end{eqnarray}
Thus we have expressed the velocity differences and electric field in terms of
magnetic field. The evolution equation for magnetic field is given by
Eq. (\ref{Faraday-I}),
\begin{equation}
\dt \BI = \eta_{\rm eff} \nabla^2 \BI.
\end{equation}
This is a diffusion equation and the comoving diffusion length, below which
magnetic fields diffuse due to the electric resistivity in a Hubble time, is
\begin{equation}
\lambda_{\rm diff} = \frac{1}{a} \sqrt{\frac{\eta_{\rm eff}}{H}}
\approx 5 \times 10^{-5} \sqrt{1 + 2 \times 10^{-4} \gamma^{-1}} ~ {\rm pc}.
\end{equation}
The recombination increases the effective electric resistivity by a factor
$(1 + 2 \times 10^{-4} \gamma^{-1})$ and then the diffusion length by
a factor $\sqrt{1 + 2 \times 10^{-4} \gamma^{-1}}$. Therefore, the evolution
of magnetic fields at cosmological scales are not practically affected by
recombination with such a large residual ionization rate
$\gamma_{\rm end} \sim 10^{-4}$. According to \cite{KT1,KT2,KT3,TCA,KT4},
magnetic fields are generated by cosmological perturbations with scales from
$100 {\rm AU}$ to the horizon scale. To preserve magnetic fields of these scales,
it is enough to have a residual ionization rate as low as $\sim 10^{-10}$.

On the other hand, the diffusion time for a fixed coherence length is,
\begin{equation}
\frac{\tau_{\rm diff}}{H^{-1}} = \frac{H}{k^2 \eta_{\rm eff}}
\approx 5 \times 10^{20}
        \left( \frac{k^{-1}}{1 {\rm Mpc}} \right)^2
        \left( 1 + 2 \times 10^{-4} \gamma^{-1} \right)^{-1}.
\label{diffuse-time}
\end{equation}
From Eqs. (\ref{dvpe-I-sol}), (\ref{dvbH-I-sol}) and (\ref{E-I-sol}), we see that
the velocity differences and electromagnetic fields decrease with the timescale
(\ref{diffuse-time}) which is much larger than cosmological timescale.
Because magnetic fields are practically constant in time, electric currents
which support them must also be constant in time. This can be seen
in Eq. (\ref{dvpe-I-sol}), and this also means that the velocity difference
between protons and electrons increases by a factor $\gamma_{\rm end} \sim 10^{-4}$.
It would be instructive to see how protons and electrons are accelerated
during the process of recombination. However, equations of motion, Eqs. (\ref{EOMp}),
(\ref{EOMe}) and (\ref{EOMH}), at the lowest order are,
\begin{eqnarray}
\dt \v
&=& \frac{e}{\mp} \EI
    - \left( \frac{e^2 \nb \eta}{\mp} + \frac{\beta}{1+\beta} \nu_p \right) \dvpeI
    - \nu_p \dvbHI - \nabla \Phi, \\
\dt \v
&=& - \frac{e}{\me} \EI
    + \left( \frac{e^2 \nb \eta}{\me} + \frac{1}{1+\beta} \nu_e \right) \dvpeI
    - \nu_e \dvbHI - \nabla \Phi, \\
\dt \v
&=& \frac{\beta}{(1+\beta)^2} \gamma (\nu_p - \nu_e) \dvpeI
    + \gamma \frac{\nu_p + \beta \nu_e}{1+\beta} \dvbHI - \nabla \Phi,
\end{eqnarray}
all of which lead to,
\begin{equation}
\dt \v = - \nabla \Phi.
\end{equation}
Thus, at this order, all components behave as a single fluid and the interactions
among the three components vanish effectively. This is exactly expected from lowest-order
tight coupling approximation. Therefore, we must go on to the next order
where deviations among the three components begin to appear.

\subsection{tight coupling approximation II}

The evolution equations of velocity differences and Maxwell equations at the next
to the lowest order in tight coupling approximation are,
\begin{eqnarray}
&& \dt \dvpeI
   = \frac{(1+\beta) e}{\me} \EII
     - \left[ \frac{(1+\beta) e^2 \nb \eta}{\me} + \frac{\beta \nu_p + \nu_e}{1+\beta} \right]
       \dvpeII
     - (\nu_p - \nu_e) \dvbHII, \label{Ohm-II} \\
&& \dt \dvbHI
   = - \frac{\beta}{(1+\beta)^2} (1+\gamma) (\nu_p - \nu_e) \dvpeII
     - \frac{\nu_p + \beta \nu_e}{1+\beta} (1+\gamma) \dvbHII,
   \label{dvbH-II} \\
&& \dt \EI = \nabla \times \BII - e \nb \dvpeII, \label{Ampere-II} \\
&& \dt \BII = - \nabla \times \EII. \label{Faraday-II}
\end{eqnarray}
Again we solve Eqs. (\ref{Ohm-II}) and (\ref{dvbH-II}) for $\dvbHII$ and $\EII$,
\begin{eqnarray}
&& \dvbHII = - \frac{\beta}{1+\beta} \frac{\nu_p - \nu_e}{\nu_p + \beta \nu_e} \dvpeII
             + \frac{\beta (\nu_p - \nu_e)}{(\nu_p + \beta \nu_e)^2 (1+\gamma)}
               \dt \dvpeI \label{dvbH-II-dvpe} \\
&& \EII = e \nb \eta_{\rm eff} \dvpeII
          + \frac{\me}{(1+\beta) e}
            \left[ 1 + \frac{\beta (\nu_p - \nu_e)^2}{(\nu_p + \beta \nu_e)^2 (1+\gamma)} \right]
            \dt \dvpeI. \label{E-II-dvpe}
\end{eqnarray}
We do not need to solve further for $\dvpeII$ to obtain effective equations
of motion for $\vp^{(I)}$, $\ve^{(I)}$ and $\vH^{(I)}$. In fact,
substituting Eqs. (\ref{dvbH-II-dvpe}) and (\ref{E-II-dvpe}) into
Eqs. (\ref{EOMp}), (\ref{EOMe}) and (\ref{EOMH}), we have,
\begin{eqnarray}
\dt \vp^{(I)}
&=& \frac{e}{\mp} \EII
    - \left( \frac{e^2 \nb \eta}{\mp} + \frac{\beta}{1+\beta} \nu_p \right) \dvpeII
    - \nu_p \dvbHII \nonumber \\
&=& \frac{\beta}{1+\beta}
    \left( 1 - \frac{1}{1+\gamma} \frac{\nu_p - \nu_e}{\nu_p + \beta \nu_e} \right)
    \dt \dvpeI \nonumber \\
&\approx&
    \frac{\beta}{1+\beta}
    \left( 1 - \frac{1}{1+\gamma} \frac{\nu_p - \nu_e}{\nu_p + \beta \nu_e} \right)
    \frac{f}{e \nb} \nabla \times \BI, \label{p-acc} \\
\dt \ve^{(I)}
&=& - \frac{e}{\me} \EII
    + \left( \frac{e^2 \nb \eta}{\me} + \frac{1}{1+\beta} \nu_e \right) \dvpeII
    - \nu_e \dvbHII \nonumber \\
&=& - \frac{1}{1+\beta}
      \left( 1 + \beta \frac{1}{1+\gamma} \frac{\nu_p - \nu_e}{\nu_p + \beta \nu_e}
      \right) \dt \dvpeI \nonumber \\
&\approx&
      - \frac{1}{1+\beta}
      \left( 1 + \beta \frac{1}{1+\gamma} \frac{\nu_p - \nu_e}{\nu_p + \beta \nu_e}
      \right)  
      \frac{f}{e \nb} \nabla \times \BI, \label{e-acc} \\
\dt \vH^{(I)}
&=& \frac{\beta}{(1+\beta)^2} \gamma (\nu_p - \nu_e) \dvpeII
    + \gamma \frac{\nu_p + \beta \nu_e}{1+\beta} \dvbHII \nonumber \\
&=& \frac{\beta}{1+\beta} \frac{\gamma}{1+\gamma}
    \frac{\nu_p - \nu_e}{\nu_p + \beta \nu_e} \dt \dvpeI \nonumber \\
&\approx&
    \frac{\beta}{1+\beta} \frac{\gamma}{1+\gamma}
    \frac{\nu_p - \nu_e}{\nu_p + \beta \nu_e}
    \frac{f}{e \nb} \nabla \times \BI. \label{H-acc}
\end{eqnarray}
In the last equality of each equation, we used,
\begin{equation}
\dt \dvpeI = \frac{1}{e \nb} \left( \eta_{\rm eff} \nabla^2 + f \right) \nabla \times \BI
\approx \frac{f}{e \nb} \nabla \times \BI,
\end{equation}
because $\eta_{\rm eff} k^2 \ll H \lesssim f$. We see that purely second order terms
have canceled and effective accelerations are expressed just by $\BI$.
The net acceleration of protons is contributed from electric fields and collision
with neutral hydrogens and they correspond to the first and second term
in $( \cdots )$ in Eq. (\ref{p-acc}), respectively. The former is dominant
in the early phase of recombination while the latter is dominant in the late
phase. The acceleration is proportional to the recombination rate $f$ and
inversely proportional to the baryon density $\nb$. Thus the acceleration of protons
increases substantially as recombination proceeds. The situation is the same
for electrons except that the first term contributed from electric fields is
always dominant. Thus we can see that there is no acceleration without recombination,
as it should be, and we could confirm that charged particles are accelerated
to compensate for the decrease in their number densities and maintain
the electric currents. It should be noted that the direction of acceleration
is always the opposite between protons and electrons and the compensation is
mainly realized by the acceleration of electrons.

For completeness, we will derive the evolution equation of magnetic fields
at this order. From Eq. (\ref{Ampere-II}),
\begin{eqnarray}
e \nb \dvpeII
&=& \nabla \times \BII - \dt \EI \nonumber \\
&=& \nabla \times \BII - \eta_{\rm eff}^2 \nabla^2 \nabla \times \BI
    - \frac{1}{\omega_p^2} \frac{\nu_p \nu_e}{\nu_p + \beta \nu_e} (1+\gamma) f
      \nabla \times \BI.
\end{eqnarray}
Substituting this into Eq. (\ref{Faraday-II}) gives the evolution equation
for magnetic fields,
\begin{eqnarray}
\dt \BII
&=& - e \nb \eta_{\rm eff} \nabla \times \dvpeII
    - \frac{\me}{(1+\beta) e}
      \left[ 1 + \frac{\beta (\nu_p - \nu_e)^2}{(\nu_p + \beta \nu_e)^2 (1+\gamma)} \right]
      \dt \nabla \times \dvpeI \nonumber \\
&=& \eta_{\rm eff} \nabla^2 \BII
    - \left[ \eta_{\rm eff}^2
             - \frac{1}{(1+\beta) \omega_p^2}
               \left\{ 1 + \frac{\beta (\nu_p - \nu_e)^2}{(\nu_p + \beta \nu_e)^2 (1+\gamma)} \right\}
      \right] \eta_{\rm eff} \nabla^4 \BI \nonumber \\
& & - \left[ \frac{1}{\omega_p^2} \frac{\nu_p \nu_e}{\nu_p + \beta \nu_e} (1+\gamma) \eta_{\rm eff}
             - \frac{1}{(1+\beta) \omega_p^2}
               \left\{ 1 + \frac{\beta (\nu_p - \nu_e)^2}{(\nu_p + \beta \nu_e)^2 (1+\gamma)} \right\}
      \right] f \nabla^2 \BI.
\end{eqnarray}
Picking up the dominant term, we finally have,
\begin{equation}
\dt \BII = \eta_{\rm eff} \nabla^2 \BII
           + \frac{f}{(1+\beta) \omega_p^2} \nabla^2 \BI.
\end{equation}

\section{Discussion and summary \label{section:discussion}}

In our analysis above, we neglected photons and let us mention a couple of potentially
important effects. One is effective electric resistivity due to Thomson scattering
between charged particles and photons. The relative importance of the effective
resistivity $\eta_{\rm Th}$ and the normal resistivity $\eta$ is expressed as
\cite{KT4},
\begin{equation}
\frac{\eta_{\rm Th}}{\eta}
= \frac{1+\beta^4}{(1+\beta)^2} \frac{\sigma_T \rho_{\gamma}}{e^2 n_b \eta}
\approx 4 \times 10^{-8} x_p^{-1} \left( \frac{1+z}{10^3} \right)^{5/2},
\end{equation}
where $\sigma_T$ is Thomson cross section and $\rho_{\gamma}$ is the photon energy density.
Thus, we can see that the effective resistivity due to Thomson scattering play a minor role
for a reasonable residual ionization rate. On the other hand, photons also affect
the dynamics of charged particles. In particular, Ohm's law will have a contribution
from Thomson scattering other than the effective resistivity. The analysis of this term
requires a four fluid treatment, which is beyond the scope of this paper.

In this paper, we followed the evolution of protons, electrons, neutral hydrogens
and magnetic fields around recombination using tight coupling approximation and
cosmological perturbation theory. We found that neutral hydrogens act as an effective
electric resistivity which depends on the ionization rate and that the residual ionization
rate required to maintain magnetic fields at cosmological scales is about $10^{-10}$,
which is much smaller than the standard value $\sim 10^{-4}$. Further, we showed
that charged particles, especially electrons, are accelerated to compensate the decrease
in their number density to maintain electric currents.

\acknowledgements

The author is grateful to K. Ichiki for useful discussion
and S. Inutsuka for helpful comments. KT is supported by
a Grant-in-Aid for the Japan Society for the Promotion of
Science Fellows and are research fellows of the Japan Society
for the Promotion of Science.

\end{document}